\documentclass[aps,prl,preprint,nopacs,superscriptaddress]{revtex4}
\usepackage{graphicx}
\usepackage{verbatim}
\usepackage{mathrsfs}
\usepackage{color}
\usepackage{amsmath,amsfonts,amssymb}
\usepackage{graphicx}
\def\3{2.8in}    %used for figure widths
\def\2{2.5in}
\def\4{3.0in}

\def \beq {\begin{equation}}
\def \eeq {\end{equation}}
\begin{document}
\renewcommand{\figurename}{\textbf{Figure}}
%\begin{CJK*}{GB}{gbsn} % Use default fonts from CJK (see below)

\title{Spinor-dominated magnetoresistance driven by the topological phase transition in $\beta $-Ag$_2$Se  }

\author{Cheng-Long Zhang$^{\ast}$$^{\dagger}$}
\affiliation{Beijing National Laboratory for Condensed Matter Physics, Institute of Physics, Chinese Academy of Sciences, Beijing 100190, China}
\affiliation{International Center for Quantum Materials, School of Physics, Peking University, Beijing 100871, China}

\author{Yilin Zhao$^{\ast}$}
\affiliation{Division of Physics and Applied Physics, School of Physical and Mathematical Sciences, Nanyang Technological University, Singapore, Singapore}

\author{Yiyuan Chen$^{\ast}$}
\affiliation{Department of Physics and Shenzhen Institute for Quantum Science and Engineering, Southern University of Science and Technology, Shenzhen 518055, China}

\author{Ziquan Lin\footnote{These authors contributed equally to this work}}
\affiliation{Wuhan National High Magnetic Field Center and School of Physics, Huazhong University of Science and Technology, Wuhan 430074, China}

\author{Sen Shao}
\affiliation{Division of Physics and Applied Physics, School of Physical and Mathematical Sciences, Nanyang Technological University, Singapore, Singapore}

\author{Zhen-Hao Gong}
\affiliation{Department of Physics and Shenzhen Institute for Quantum Science and Engineering, Southern University of Science and Technology, Shenzhen 518055, China}

\author{Junfeng Wang$^{\dagger}$}
\affiliation{Wuhan National High Magnetic Field Center and School of Physics, Huazhong University of Science and Technology, Wuhan 430074, China}

\author{Hai-Zhou Lu$^{\dagger}$}
\affiliation{Department of Physics and Shenzhen Institute for Quantum Science and Engineering, Southern University of Science and Technology, Shenzhen 518055, China}

\author{Guoqing Chang\footnote{Corresponding authors: chenglong.zhang@iphy.ac.cn, luhz@sustech.edu.cn, guoqing.chang@ntu.edu.sg, jfwang@hust.edu.cn}}
\affiliation{Division of Physics and Applied Physics, School of Physical and Mathematical Sciences, Nanyang Technological University, Singapore, Singapore}

\author{Shuang Jia}
\affiliation{International Center for Quantum Materials, School of Physics, Peking University, Beijing 100871, China}
\affiliation{Interdisciplinary Institute of Light-Element Quantum Materials and Research Center for Light-Element Advanced Materials, Peking
University, Beijing 100871, China}

\begin{abstract}
A topological insulator is a quantum material which possesses conducting surfaces and an insulating bulk. Despite extensive researches on the properties of Dirac surface states, the characteristics of bulk states have remained largely unexplored. Here we report the observation of spinor-dominated magnetoresistance anomalies in the topological insulator $\beta $-Ag$_2$Se, induced by a magnetic-field-driven band topological phase transition. These anomalies are caused by intrinsic orthogonality in the wave-function spinors of the last Landau bands of the bulk states, in which backscattering is strictly forbidden during a band topological phase transition. This new type of longitudinal magnetoresistance, purely controlled by the wave-function spinors of the last Landau bands, highlights a unique signature of electrical transport around the band topological phase transition. With further reducing the quantum limit and gap size in $\beta $-Ag$_2$Se, our results may also suggest possible device applications based on this spinor-dominated mechanism and signify a rare case where topology enters the realm of magnetoresistance control.

\end{abstract}

\maketitle
%\linenumbers
A 3-dimensional (3D) topological insulator (TI) is characterized by insulating bulk and surface Dirac cone states \cite{hasan2010colloquium, qi2011topological}. The surface Dirac cone states have been extensively studied \cite{hsieh2008topological, qu2010quantum, analytis2010two, chang2013experimental}, but the bulk bands have received little attention. The inverted bulk bands in a TI are nontrivial \cite{zhang2009topological, liu2010model}, and can undergo a  band topological phase transition (TPT) with a gap closing point as the quantum critical point. The transport properties near the band TPT remain largely unexplored \cite{BJY, BJY_NP}, unlike the non-Fermi-liquid behaviors near the quantum critical point in strongly correlated materials \cite{QPT},

The band TPT in nonmagnetic TIs can be triggered by changes in chemical composition, pressure, strain or external magnetic field \cite{murakami2007phase}. When the band gap is small and a magnetic field is present, a unique band TPT occurs on Landau bands (LBs). Spectroscopic measurements have mainly explored these band TPTs \cite{xu2012observation, liang2017, zhang2021berry, JHChu, chen2017spectroscopic, wu2023topological}, but they cannot reveal the exotic physical properties near the band closing point. To our knowledge, no distinct transport signatures have been uncovered on the transition point. However, electrical transport under a strong magnetic field provides a unique opportunity to explore the physics near the gap closing point by continuously going through the band TPT. In this study, we report a significant transport signal in $\beta $-Ag$_2$Se, a band TPT-induced spinor-dominated magnetoresistance (MR) anomaly. During a magnetic-field-driven band TPT, the presence of two intrinsically backscattering forbidden 1-dimensional (1D) conducting channels in a 3D TI, mimics the ballistic helical edge modes in 2-dimensional (2D) TIs.

Before delving into the experimental results, let us first discuss the unique physics behind the band TPT of LBs in TI. In a TI, the bulk band splits into a series of LBs with strong orbital quantization under a magnetic field. The energies of the LBs are determined by two key scales, Zeeman splitting ($g^*\mu_{B}$B) and orbital cyclotron energy ($\hbar\omega_c$), where $\hbar$ is the Planck constant divided by $2\pi$, $\omega_c$=$e$B/$m^*$ is the cyclotron frequency, $e$ is the elementary charge of electron, $m^*$ is the effective mass, $g^*$ is the effective $g$-factor, and $\mu_{B}$ is the Bohr magneton. For a free electron system, $\hbar\omega_c$ is equal to $g^*\mu_{B}$B, while $\hbar\omega_c$ dominates in topological materials with ultra-light $m^*$ ($\sim$ 0.1$m_0$), here $m_0$ is the bare electron mass. Due to the band inversion of TI, the orbital cyclotron and Zeeman effects will disentangle the two inverted bands, leading to a magnetic-field-driven band TPT as illustrated in Fig. 1a. The Zeeman effect can be treated as a modified orbital cyclotron effect due to the spin-polarized nature of LLBs, sharing the same linear-in-B relation with orbital cyclotron effect. We can describe this process in details by using a typical $\boldsymbol{k}$$\cdot$$\boldsymbol{p}$ model of TI, written as $H(\vec{k})=M_{\mathbf{k}}\tau_z+\hbar(\sum\limits_{a=x,y,z}v_ak_a\Gamma_a)$, where $\tau_z$ labels for the orbitals while the second term is the massless Dirac equation based on the $\Gamma$ matrices. This model is characterized by a mass control term $M_{\mathbf{k}}=M_0+M_\bot(k_y^2+k_z^2)+M_xk_x^2$, where $M_0$, $M_x$ and $M_\perp$ are band parameters for band gap at $k_x=0$, in- and out-of-plane parabolic energy dispersions, respectively \cite{zhang2009topological}. In the ultraquantum limit with B $\parallel$ $\textbf{$x$}$ (\textbf{a})-axis, there are only one pair of spin-polarized last LBs (LLBs) on the Fermi level. The energy dispersion of the two LLBs is $E_0$$_\pm$ = $\sqrt{m^2+\hbar^2k_x^2v_x^2}$, where the mass term (or equivalently, the band gap at $k_x=0$) of the two LLBs is $m$ = $M_0$ + $M_xk_x^2$ + $M_\perp/l_B^2$, where $l_B=\sqrt{\hbar/eB}$ is the magnetic length, controlling the cyclotron energy. At $k_x=0$, the inverted band is directly controlled by the mass term $m$ = $M_0$ + $M_\perp/l_B^2$, where signs of $M_0$ and $M_\perp$ are opposite due to the inverted dispersion. Then, linear-in-B cyclotron energy will lever the inverted gap $M_0$ and finally dominates by crossing a gapless intermediate state with $m$ = 0, underpinning the process of the band TPT as illustrated in Fig. 1a. Furthermore, we can see that the critical field $\mu_0H^c$ for $m$ = 0 is determined by $M_0$/$M_\perp$, causing $\mu_0H^c$  proportional to 1/$M_\perp$, which can be reflected by the effective mass $m^*$ in transport measurements. Then, anisotropic $m^*$ measured from an ellipsoid-like Fermi surface can roughly track $\mu_0H^c$, namely $m^*$ $\propto$ $\mu_0H^c$.

The consequence of $m$ = 0 can be envisioned by looking at the highly nontrivial properties of LLB whose wave function uniquely contains a spinor eigenvector, inherited from the 4$\times$4 matrices-based Dirac Hamiltonian and a normal part from the harmonic oscillator. As shown in Fig. 1b, the two spinors of +$k_x$ and -$k_x$ become orthogonal when $m$ = 0 at the band TPT, here the $M_xk_x^2$ in the mass term $m$ is treated to be negligible in the ultraquantum limit due to the field suppression of Fermi level. Consequently, as shown in Fig. 1b, the LLBs become gapless at a critical field ($\mu_0H^c$), causing the transport form factor $I_s$ which is proportional to the transition matrix between the two spinors at +$k_x$ and -$k_x$ to vanish identically \cite{chen2018forbidden}.
The vanished $I_s$ leads to an extremely long transport relaxation time ($\tau\propto1/I_s$), making the backscattering forbidden (Fig. 1b) and leading to a pronounced dip in longitudinal MR at $\mu_0H^c$ (Fig. 1c). Because the 1D dispersion of LLB is always determined by the direction of magnetic field, this additional transport channel occurs only when the magnetic field is along the current direction, making the longitudinal MR a unique probe to detect this anomaly. The longitudinal MR anomalies here differ from the forbidden backscattering in graphene \cite{ando1998berry}, where electrons on the zero-field Dirac band, instead of LBs, are forbidden from backscattering due to the natural orthogonality in Pauli matrices. The anomaly also differs from the chiral anomaly, a 1+1D chiral fermion effect with unbalanced left and right movers on any 1D band dispersions across the Fermi level \cite{witten2016three}. However, the forbidden backscattering may occur twice at respective critical fields depends on the relative sign between $M_0$, $M_x$ and $M_\perp$ (see detailed discussions on Fig. 4), it is absent in a Weyl band because the scattering matrix of LLBs' spinors cannot be zero \cite{chen2018forbidden}.

To observe this unique longitudinal MR, we require a TI with only one pair of inverted bands around the Fermi level and a narrow gap accessible by a magnetic field. Unfortunately, narrow-gapped strong TIs are rare, and most contain complicated band structures near the Fermi level \cite{akimov1993carrier, assaf2017negative}. In this work, we have chosen $\beta $-Ag$_2$Se \cite{zhang2017ultraquantum, zhang2011topological}, as a 3D strong TI, an ideal platform to achieve the spinor-dominated MR. As shown in Fig. 1d and 1e, $\beta $-Ag$_2$Se, appears as long ribbons and crystallizes in an orthorhombic unit cell. The electronic band calculation by Heyd-Scuseria-Ernzerhof (HSE) approximation indicates that $\beta $-Ag$_2$Se is a strong TI with $Z_2$ = (1; 0 0 0) and an indirect band gap around 4$\sim$10 meV (Fig. 1f and 1g). This is compared with the Perdew-Burke-Ernzerhof (PBE) approximation, which usually underestimates the gap and shows that $\beta $-Ag$_2$Se is a semimetal.

We conducted low-field electrical transport measurements on sample S1 with carrier concentration $n$ = 1.1$\times$10$^{18} $ cm$^{-3}$, and as shown in Fig. 1h and 1i, we observed clear single-frequency Shubnikov-de Haas (SdH) oscillations with a quantum limit around 3 T when the magnetic field is tilted towards $\textbf{b}$($\theta$) and $\textbf{c}$($\phi$) axes. We found no new frequencies in all the mapped ranges of angles. The angular dependence of extremal cross-sectional areas, with frequencies $F_{\theta, \phi}$, depicts a 3D anisotropic Fermi pocket of the bulk state consistent with the calculation (see section I of SI for details). The small electron pocket and narrow band gap make $\beta $-Ag$_2$Se as an ideal platform for investigating the physics of pure inter-1D channels (LLBs) scattering in the ultra-quantum limit under a modestly strong magnetic field. We then conducted electrical transport measurements on another sample S2 in a strong magnetic field because it has a lower quantum limit $\sim$ 2 T. Figure 2a shows the complicated dependence of longitudinal MR ($\Delta\rho_{xx}(H)/\rho_0$) at 1.5 K characterized by two anomalous dips at 1.5 K, denoted as B1 at 5 T and B2 at 51 T, with the current and magnetic field applied along the same crystallographic \textbf{a} direction.

To investigate the physical mechanism underlying B1 and B2, we tracked the two anomalies by performing temperature and angle dependence of $\Delta\rho_{xx}(H)/\rho_0$. Figure 2b shows B1 and B2 gradually vanish as temperature increases, results in a smooth background that persists even at 70 K. The critical field of B2 slightly shifts to a lower field when temperature rises, which is against many-body effects such as a charge density wave (CDW) transition in a magnetic field  \cite{fauque2013two}.
Moreover, the angular dependence of $\Delta\rho_{xx}(H)/\rho_0$, as shown in Fig. 2c, shows that the B1 and B2 shift towards higher fields when the direction of the magnetic field is tilted away from the current direction, indicating that B2 is affected by the anisotropy of band dispersion. The temperature dependence of B2 suggests a specific band effect that causes the MR dips, rather than a many-body effect.

We attempted to understand the anomalies B1 and B2 in the ultraquantum limit by examining the relationship between the smooth background of $\Delta\rho_{xx}(H)/\rho_0$ and B1, B2. Figure 3a shows conductance $G_{xx}$ at each tilted angle ($\theta$), defined as $1/\rho_{xx}(H)$, against the out-of-plane magnetic field component. We observed that the backgrounds of $G_{xx}$ align on a single curve at each tilted angle, indicating a normal orbital MR effect. Additionally, the peak values of $G_{xx}$ replotted against $\theta$ in Fig. 3b can be fitted by a cos$^n\theta$ ($n$ = 4.6) empirical relation.
According to the Shockley-Chambers formula \cite{ziman2001electrons}, the angle-dependent conductivity of a cylinder-shaped Fermi surface yields a $\sim$cos$^2\theta$ relation due to the anisotropy of $m^*$. If the transport relaxation time $\tau$ is isotropic, the angle-dependent conductivity dampens slower than a $\sim$cos$^2\theta$ relation for an anisotropic Fermi surface. The obvious deviation on exponent $n$ (Fig. 3b) shows that the isotropic transport relaxation time $\tau$ assumption must break down, indicating the existence of an additional conducting channel only when the magnetic field is tilted along the direction of the current.

We further explored this additional conducting channel by examining the related control parameters. As shown in Fig. 3c, the fields of B2, denoted as $\mu_0H^c_{B2}$, linearly increases with rising temperature, indicating that the thermal energy ($k_BT$) competes with cyclotron/Zeeman energy which is proportional to the magnetic field. On the other hand, $\mu_0H^c_{B1}$ and $\mu_0H^c_{B2}$ follow the same angular dependence of SdH oscillations at 4.2 K as shown in Fig. 1i. This angular dependence is obvious when we scaled $\mu_0H^c_{B1,2}(\theta)/\mu_0H^c_{B1,2}(0^o)$ versus $F_\theta$/$F_{\theta=0^o}$ in Fig. 3e. The frequency of SdH oscillations is expressed by the quantization rule $F=\frac{\hbar}{2\pi{e}}\mathrm{A_F}$, where $e$ is the elementary charge, and $\mathrm{A_F}$ is the extremal cross-sectional area, which is roughly proportional to the $k_F^2$. Then, $F$ reflects the anisotropy of effective mass ($m^*$) if energy dispersion $\epsilon=\frac{\hbar^2k_F^2}{2m^*}$ is adopted, while $m^*$ directly determines the cyclotron energy $\hbar\omega_c$. As we already pointed out in the introductory part, the $m^*$ is proportional to the critical field $\mu_0H^c_{B2}$ of the band TPT on the LLBs, our experimental observation of the coincidence between $\mu_0H^c_{B2}$ and $m^*$ confirms the longitudinal MR anomaly is caused by a band TPT-induced mechanism.

We ascribe the anomaly in longitudinal MR induced by the band TPT to intrinsic forbidden backscattering on LLBs caused by spinor orthogonality around the phase transition, as illustrated in Fig. 1. To support our interpretation, we conducted a detailed simulation on $\beta$-Ag$_2$Se. Before quantitative simulation, we thoroughly inspect the mass term of LLBs $m$ = $M_0$ + $M_xk_x^2$ + $M_\perp/l_B^2$. The magnetic dependence of this term is important, because it controls the form factor $I_s$ ($\propto m^2$). In this well-established TI model, the relative signs of $M_0$, $M_x$ and $M_\perp$ determine the band inversions along different high-symmetric $\bf{k}$ paths. $M_0$$M_x$$<$0 and $M_0$$M_\perp$$<$0 represent a strong TI, while only one of them satisfied represents a weak TI, and none of them satisfied represents a normal insulator \cite{chen2018forbidden}. We then simulated the magnetic field dependence of $m$ in different topological phases. As shown in Fig. 4a-c, $m$ ($I_s$) shows no zero crossing (no dip), one zero crossing (one dip) and two zero crossings (two dips) in the quantum limit of normal insulator, weak TI and strong TI, respectively. We found that the field dependence of $m$ exhibits distinct behaviors in the three phases based on the detailed mathematical structures of $m$, which depends on the relative signs of $M_0$, $M_x$ and $M_\perp$. Therefore, we can expect different numbers of longitudinal MR dips in different topological phases when $m$ vanishes across the band TPT.

By adopting a $\boldsymbol{k}$$\cdot$$\boldsymbol{p}$ model with specific parameters from band calculations (refer to section III in SI), our aim is to determine the values of $\mu_0H^c$ where anomalies occur. The main parameter is the mass control term $m$ = $M_0$ + $M_xk_x^2$ + $M_\perp/l_B^2$ with $M_0$ = -0.036 eV fixed by direct energy difference around $\Gamma$ point. If we ignore the Zeeman energy, we map out the zero crossings at the critical magnetic field, $\mu_0H^c_{B2}$ versus $M_x$ and ${M}_\perp$ in Fig. 4d. As mentioned before, the $M_xk_x^2$ term contracts into a negligible value in a strong magnetic field. Therefore, we can see that $\mu_0H^c_{B2}$ is sensitively dependent on ${M}_\perp$, and the value from band fitting (red dotted curve) indicates $\mu_0H^c_{B2}$ around 180 T, which contradicts our experimental B2 anomaly around 51 T. While the above estimation is solely based on the cyclotron energy ($M_\perp/l_B^2$) of LLBs without including the Zeeman effect. As we discussed before, the Zeeman effect on LLBs can be easily included as a modified cyclotron energy, then the mass term can be written as $m$ = $M_0$ + $M_xk_x^2$ + $\tilde{M}_\perp/l_B^2$, where $\tilde{M}_\perp=M_\perp+g^*\mu_B\hbar/4e$ is the modified in-plane mass parameter. As shown in Fig. 4e, $\mu_0H^c_{B2}$ decreases when $g^*$ increases. The bold pink curve indicates the predicted $\mu_0H^c_{B2}$ based on $g^*$ from band calculations, where a range of $g^*$ is used for a reasonable comparison with the experimental B2 anomaly due to the anisotropy of $\beta $-Ag$_2$Se (see section II of SI for details on calculations of $g$-factors). We can now see that the $\mu_0H^c_{B2}$ of anomaly B2 falls within the simulated range, which supports the spinor-dominated mechanism driven by band TPT. The remaining anomaly $\mu_0H^c_{B1}$ cannot be precisely determined by simulation due to large uncertainty in the estimation of $M_xk_x^2$ term as explained before. Despite this, the existence of two anomalies is consistent with the strong TI phase in $\beta$-Ag$_2$Se.

In conclusion, we have observed large longitudinal MR dips in the ultraquantum limit of the strong TI $\beta $-Ag$_2$Se. The ultraquantum-limit anomaly points to an underlying mechanism based on spinor-dominated forbidden backscattering on LLBs, driven by band TPT. The magnetic field acts as a tunable knob that controls the access of distinct topological phases. Therefore, the newly discovered MR effect presents a rare MR effect, with a pure topological origin, that can be adopted to design logic devices with on/off functions. Our study also signifies the physics comes from bulk bands of TI, mainly connected with band TPT, where a plethora of unexplored physics should be targeted in the future.

\vspace {2cm}

% ------------------------------------------------------------------------
\bibliographystyle{unsrt}

\clearpage

\textbf{Methods}

\textbf{Sample preparation and characterization.} Single crystals of $\beta$-Ag$_2$Se can be grown by modified vapor transfer method as described in ref.\cite{zhang2017ultraquantum}. Polycrystalline $\beta$-Ag$_2$Se was sealed in a tube silica ampoule, and then placed in a tube furnace subjected to a temperature gradient from 500$^o$C to around room temperature for several days. The shape of as-grown samples are ribbon-like, and the crystalline axes were determined by single crystalline X-Ray diffraction measurements as described in ref.\cite{zhang2017ultraquantum}.

\textbf{Transport measurements.} Magneto-transport measurements (56 T) were performed at the Wuhan National High Magnetic Field Center (WHMFC). A Digital lock-in technique was employed with $f$ = 100 kHz and $i$ = 5 mA by using a non-destructive pulse magnet with a pulsed duration of $60$ ms. Resistance was measured by a standard 4-probe method. Measurements with both positive and negative field polarities were performed to eliminate the effects of contact asymmetries. Data for the up-sweeping and down-sweeping of the pulse field were in good agreement, thus the self-heating effect of the sample, in the pulsed high magnetic fields, can be excluded.

\textbf{First-principles calculations.} First-principles calculations of $\beta$-Ag$_2$Se were performed using density functional theory implemented in the Vienna ab initio simulation package (VASP) code \cite{kresse1996efficient}. The energy cutoff for plane wave basis was set at 400eV. The Brillouin zone was sampled within the Monkhorst-Pack scheme and the k mesh was set as $8\times5\times4$ . The generalized gradient approximation (GGA) expressed by the Perdew-Burke-Ernzerhof (PBE) functional was implemented for the exchange-correlation energy \cite{perdew1996generalized}. To achieve a better description of electronic properties and band gaps, screened hybrid functional of Heyd-Scuseria-Ernzerhof (HSE06) was used in band structure calculations \cite{heyd2003hybrid} with $25\%$ of the nonlocal Fock exchange combining with $75\%$ of the PBE exchange. Hartree-Fock screening was set to be 0.2 to achieve a good balance between accuracy and computational cost.

\textbf{Acknowledgments}

C.-L. Zhang thanks Junyi Zhang for the helpful discussions on the additional relaxation channel in Weyl semimetal, which initiated the thinking along this line. We thank Titus Neupert for discussions and comments on the data. C.-L. Zhang was supported by a start-up grant from IOP, CAS. S.J. was supported by the National Key Research and Development Program of China (2021YFA1401902) and the National Natural Science Foundation of China No. 12141002, No. 12225401. Work at Nanyang Technological University  was supported by the National Research Foundation, Singapore under its Fellowship Award (NRF-NRFF13-2021-0010) and the Nanyang Technological University startup grant (NTUSUG). H.-Z. Lu was supported by the National Key R\&D Program of China (2022YFA1403700), the National Natural Science Foundation of China (11925402), Guangdong province (2020KCXTD001 and 2016ZT06D348), the Science, Technology and Innovation Commission of Shenzhen Municipality (ZDSYS20170303165926217, JAY20170412152620376, and KYTDPT20181011104202253). The numerical calculations were supported by Center for Computational Science and Engineering of SUSTech. J.W. was supported by NSFC NO. U1832214, 12074135.

\textbf{Author contributions}

C.-L.Z. and S.J. conceived and designed the experiment. C.-L.Z. and Z.L. performed all transport experiments with help from J.W.; C.-L.Z. grew the single crystals; Y.Z., S.S. and G.C. performed first-principles band calculations; Y.C., C.-L.Z., Z.-H.G. and H.-Z.L. did the theoretical analyses and simulations; C.-L.Z. wrote the paper with inputs from all authors. C.-L.Z. and S.J. were responsible for the overall direction, planning and integration among different research units.

\textbf{Additional information}

Supplementary information is available in the online version of the paper. Reprints and
permissions information is available online at xxxxxxxxxx.
Correspondence and requests for materials should be addressed to C.-L.Z., H.-Z.L., G.C. or J.F.W..

\textbf{Competing financial interests}

The authors declare no competing financial interests.

\textbf{Data availability.}

The data that support the plots within this paper and other findings of this study are available from the corresponding author upon reasonable request.

\clearpage

\begin{figure}
  \includegraphics[clip, width=1\textwidth]{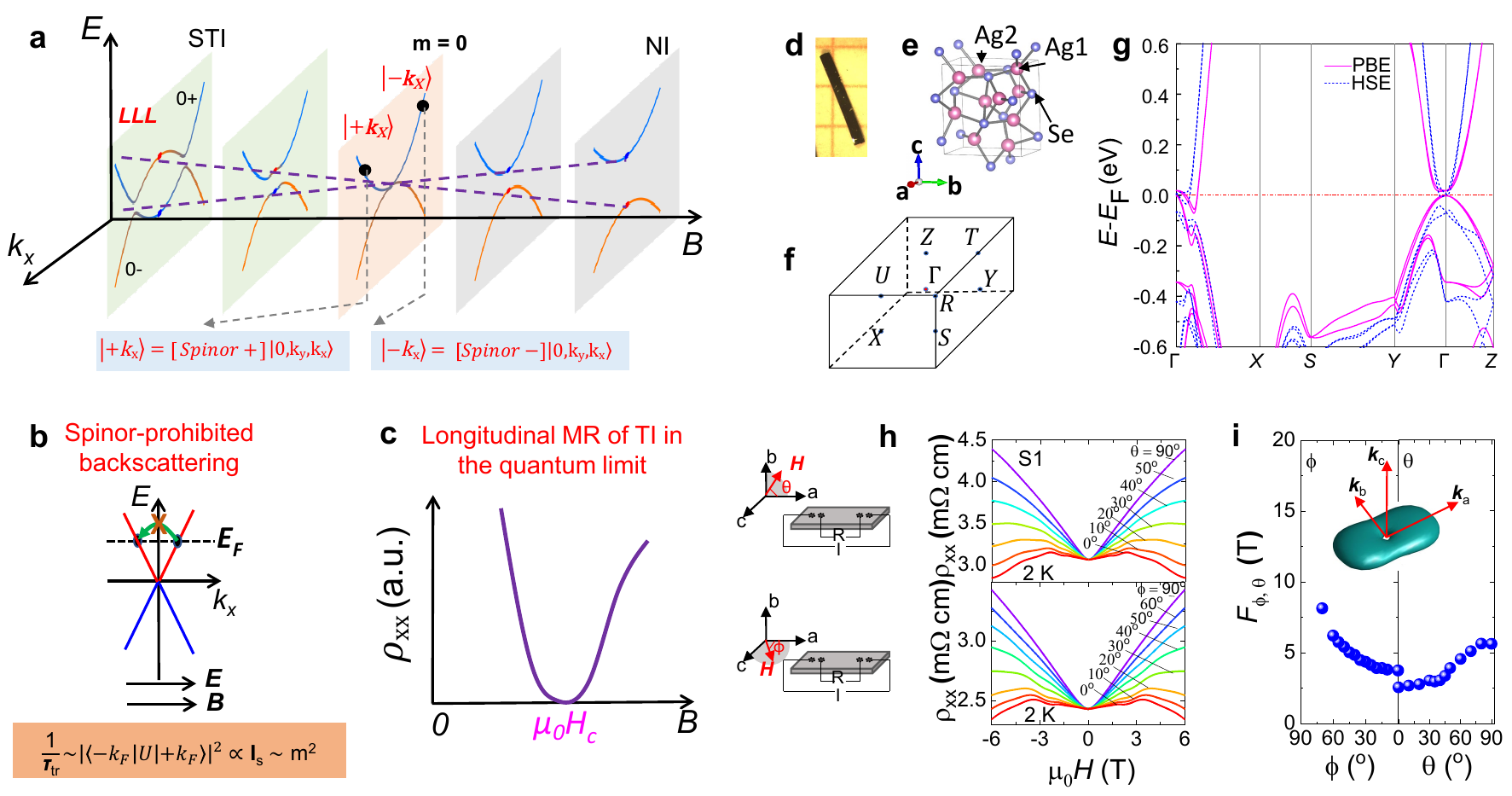}\\
  \caption{ \textbf{Topological-phase-transition-induced forbidden backscattering and band properties of $\beta $-Ag$_2$Se.} (a) The two last Landau bands (LLBs) of typical strong topological insulator (TI) go across a phase transition driven by external magnetic field. Only two states, denoted as black dots $|\pm{k_x}\rangle$, are involved in this 1D channel. (b) $|\pm{k_x}\rangle$ are orthogonal to each other when mass term $m$ is zero, which causes the spinor-prohibited backscattering. (c) As a result, a dip appears on the longitudinal magneto-resistance (MR) at critical magnetic field (B$_c$). (d) A photo of single crystalline $\beta $-Ag$_2$Se. (e) The orthorhombic crystal structure of $\beta $-Ag$_2$Se. (f) High symmetric \textbf{k} point in a unit cell. (g) Band calculations of $\beta $-Ag$_2$Se based on PBE and HSE approxiamations. (h) Angle-dependent Shubnikov-de Haas (SdH) oscillations in \textbf{ab} ($\theta$) and \textbf{ac} ($\phi$) planes. (i) The extracted frequencies of SdH along the two tilted angles, $\theta$ and $\phi$, respectively. Inset shows the mapped Fermi surface with a Fermi level determined by experimental carrier concentration ($n$). The slight mismatch of frequency around zero angle was probably caused by sample bending, in the different rounds of measurement. }
  \label{Fig1}
\end{figure}

\begin{figure}
  \includegraphics[clip, width=1\textwidth]{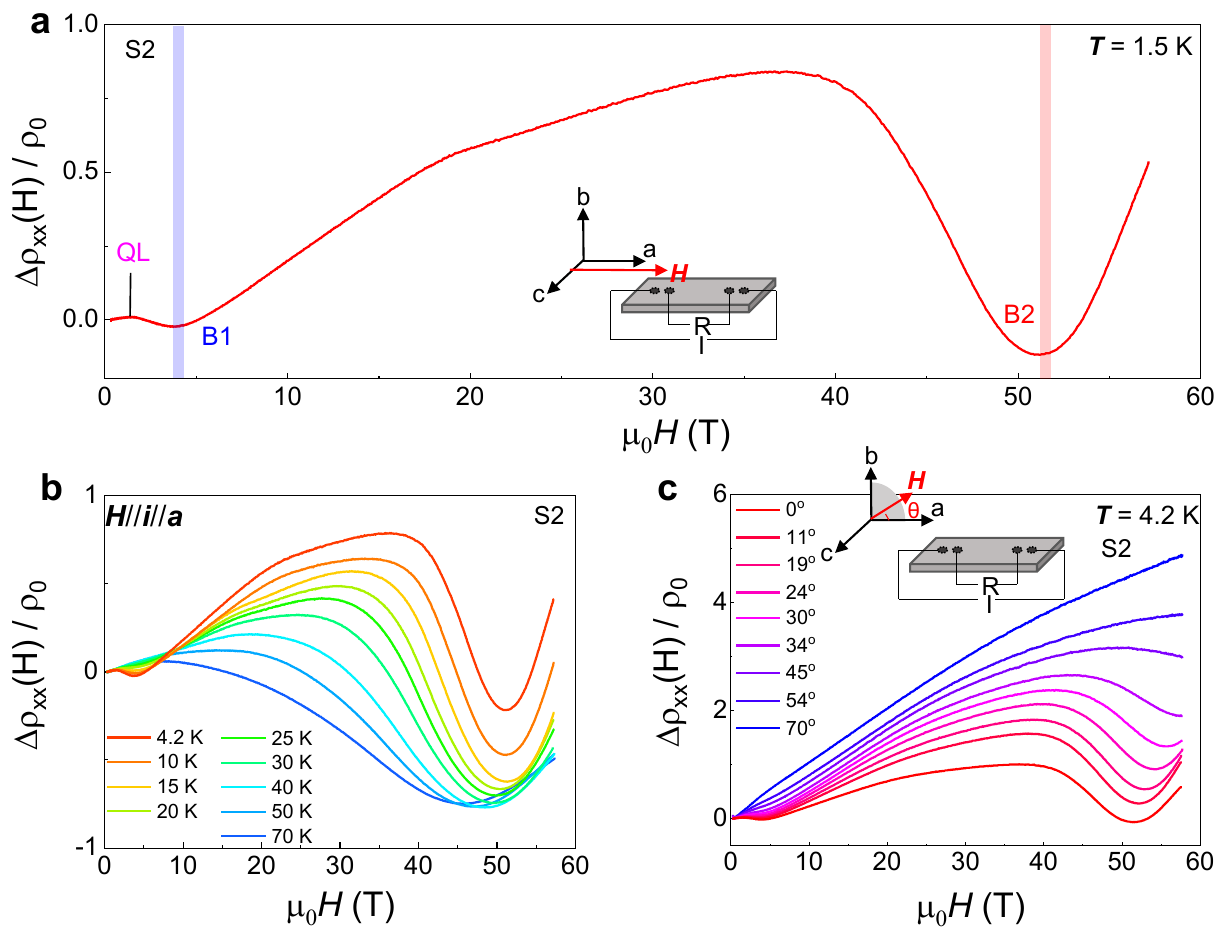}\\
  \caption{ \textbf{Ultraquantum-limit longitudinal MR of $\beta $-Ag$_2$Se.} (a) The longitudinal MR , at $T$ = 1.5 K, exhibits two anomalies denoted as B1 and B2. Inset is a sketch shows the experimental configuration of magnetic field and current. (b) Rescaled temperature-dependent  $\Delta\rho_{xx}/\rho_0$ is plotted against the magnetic field up to 56 T. (c) Rescaled angle-dependent $\Delta\rho_{xx}/\rho_0$ is plotted against the magnetic field up to 56 T. Inset shows the setup of sample rotation. }
  \label{Fig2}
\end{figure}

\begin{figure}
  \includegraphics[clip, width=1\textwidth]{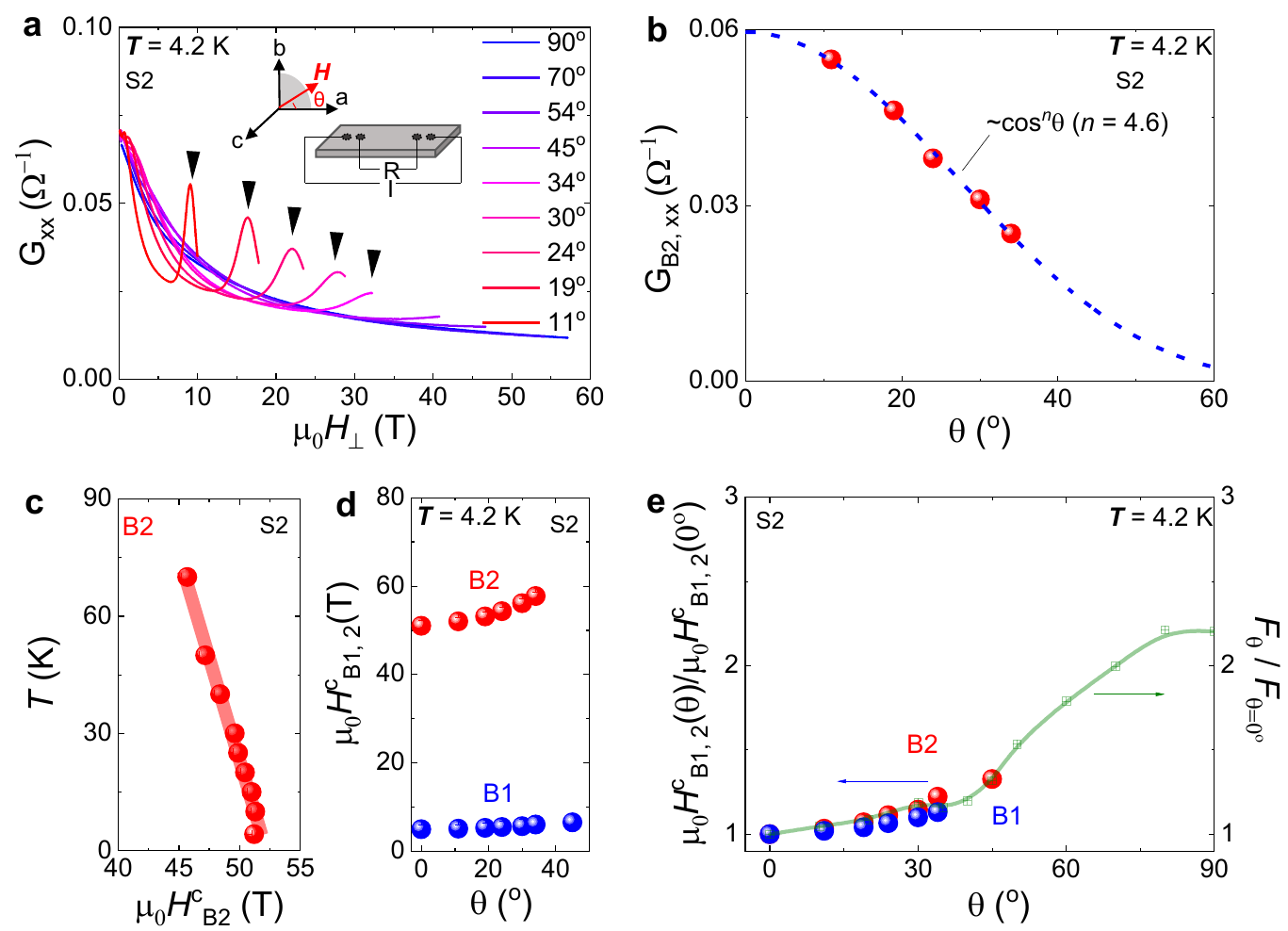}\\
 \caption{\textbf{Existence of an additional conducting channel originates from a pure band effect in $\beta $-Ag$_2$Se.} (a) The longitudinal conductance, G$_{xx}$ , rescaled versus out-of-plane component of magnetic field. (b) The amplitudes of anomaly B2 are plotted against tilted angle $\theta$, and fitted by cos$^n\theta$ function with index $n$ = 4.6, indicates the fast damping of B2 when the magnetic field is tilted away from \textbf{a}-axis. (c) The temperature-dependent critical fields of B2 shows linear T behaviors, as indicated by the linearly fitted bold red curve. (d) The angle-dependent critical magnetic fields of anomalies B1 and B2. (e) The rescaled angle dependence of B1 and B2, as shown in (d), coincides with the angle-dependent frequency of SdH oscillations, which points to the $m^*$ in cyclotron energy $\hbar\omega_c$.}
  \label{Fig3}
\end{figure}

\begin{figure}
  \includegraphics[clip, width=1\textwidth]{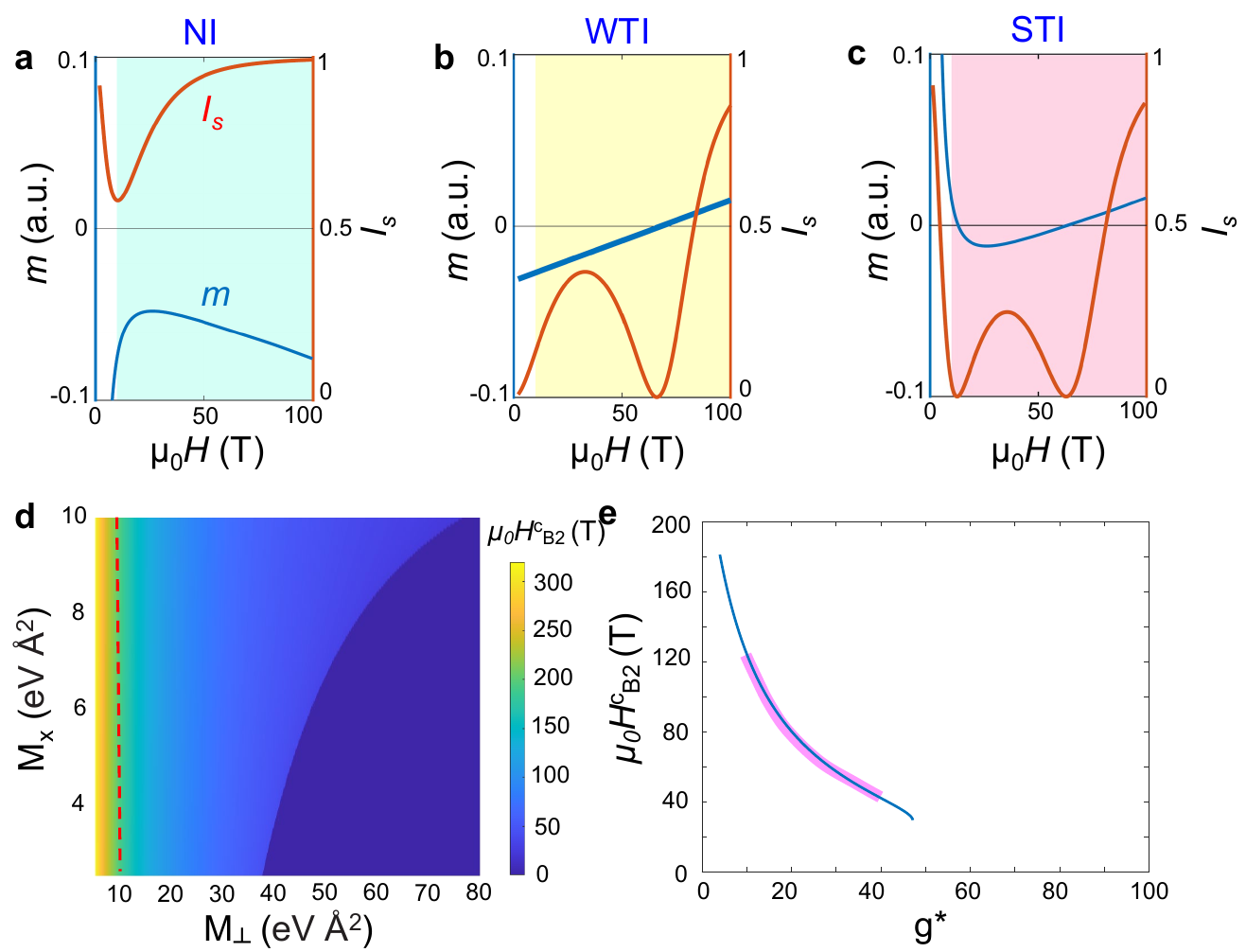}\\
  \caption{ \textbf{Simulations of the spinor-dominated longitudinal MR of in $\beta $-Ag$_2$Se based on band structure calculations.}  (a)-(c) Mass term $m$ (form factor $I_s$) shows no zero crossing (no dip), one zero crossings (one dip) and two zero crossings (two dips) in the quantum limit for normal insulator, weak TI and strong TI, respectively. (d) Color map of critical magnetic fields, B$_c$, versus coefficients M$_x$ and M$_\perp$ of a typical strong TI $\boldsymbol{k}$$\cdot$$\boldsymbol{p}$ model. Here M$_0$ is fixed to be -0.036 eV, while the dashed red line indicates the fitted value of M$_\perp$ based on band structure. (e) With considering the Zeeman effect, $\mu_0H^c_{B2}$ locates in a range, indicated by pink thick curve, which is consistent with our experimental values.     }
  \label{Fig4}
\end{figure}

\end{document}